\documentclass{PoS}

\title{Gauge invariance and gluon poles for
direct photon production}

\ShortTitle{Gauge invariance and gluon poles for
direct photon production}

\author{\speaker{I.V. Anikin}\\
        Bogoliubov Lab. of Theoretical Physics, JINR, 141980 Dubna, Russia\\
        E-mail: \email{anikin@theor.jinr.ru}}

\author{O.V. Teryaev\\
        Bogoliubov Lab. of Theoretical Physics, JINR, 141980 Dubna, Russia\\
        E-mail: \email{teryaev@theor.jinr.ru}}

\abstract{We discuss the hadron tensor of the direct photon production.
We study the effects which lead to the soft breaking of factorization
by inspection of the corresponding QCD gauge invariance.
We emphasize that the special role is played by the contour gauge for gluon fields.
We demonstrate that the different prescriptions in the gluonic pole contributions
are dictated by the presence of initial or final state interactions in diagrams.
Moreover, the different prescriptions that correspond to the
initial of final state interactions are needed to ensure the QCD gauge invariance.}

\FullConference{XXII International Baldin Seminar on High Energy Physics Problems \\
                 15-20 September, 2014\\
                 JINR, Dubna, Russia}

\begin{document}

\section{Introduction}

As shown in  \cite{AT-10}, to ensure the QED gauge invariance of
the transverse polarized Drell-Yan (DY) hadron tensor
it is a must to include
a contribution of the extra diagram which arises
from the non-trivial
imaginary part of the corresponding twist $3$ function $B^V(x_1,x_2)$.
Previously, however, this function assumed to be a real function
(see, for example, \cite{BQ} where, nevertheless,
the needed imaginary part was generated by the specially introduced ``propagator''
in the hard part of the hadron tensor).
As explained in \cite{AT-10}, the complex prescription in the representation of
$B^V$-function can be understood with a help of the contour gauge.
Moreover, the account for this extra contribution, owing to the complex $B^V$-function,
led to the amplification of the {\it hadron tensor} by the factor of $2$.
Notice that this our finding was independently confirmed in \cite{Ratcliffe13} by
using of the different approach.
The corresponding SSAs, related to the use of the twist $3$ $B^V$-function in the DY process, and the role of gluon poles were previously
discussed by many groups (see, for example,  \cite{P-R, Carlitz:1992fv, Brandenburg:1995pk, Bakulev:2007ej, Radyushkin:2009zg,
Polyakov:2009je, Mikhailov:2009sa, Brandenburg:1994wf, Teryaev, Boer, Ter00,
Ratcliffe:2009pp, Cao:2009rq, Zhou:2009jm, Ma:2003ut}).

We now present our approach, that was used in \cite{AT-10} and recently developed in \cite{Anikin:2015xka}, 
to study
the effects in the direct photon production (DPP) which lead to the soft breaking of factorization
(or the universality breaking)
by inspection of the QCD gauge invariance.
As in \cite{AT-10}, the special role is played by the contour gauge for gluon fields.
We, first, demonstrate that the prescriptions for the gluonic poles in the twist $3$ correlators
are dictated by the prescriptions in the corresponding hard parts and, second,
argue that the different prescriptions in the gluonic poles defined by
the initial or final state interactions in the diagrams under consideration (see, \cite{Anikin:2015xka} for more details).
Moreover, the different prescriptions in the representations of $B^V$-functions
are needed to ensure the QCD gauge invariance.
The situation when we have no the universality condition for the corresponding
soft function will be treated a {\it soft breaking} of factorization.

\section{Kinematics}

\begin{figure}[t]
\centerline{
\includegraphics[width=0.3\textwidth]{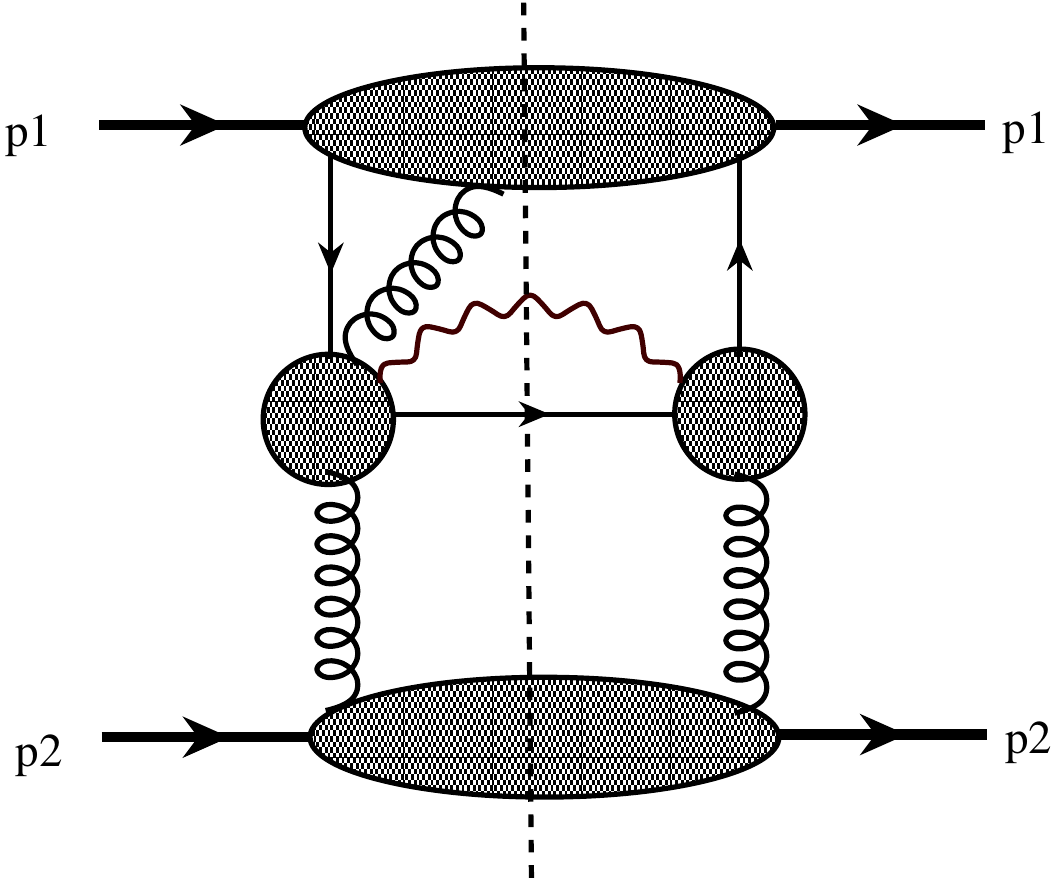}}
\caption{The Feynman diagram describing the hadron tensor of the
direct photon production.}
\label{Fig-DirPhot}
\end{figure}

We study the semi-inclusive process where the hadron with the transverse polarization
collides with the other unpolarized hadron to produce 
the direct photon in the final state in:
\begin{eqnarray}
\label{process}
N^{(\uparrow\downarrow)}(p_1) + N(p_2) \to \gamma(q) + q(k) + X(P_X)\,.
\end{eqnarray}
For (\ref{process}) (also for the Drell-Yan process), the gluonic poles manifest \cite{Teryaev}.
We perform our calculations within a {\it collinear} factorization and, therefore,
it is convenient (see, e.g., \cite{An}) to fix the dominant light-cone directions as
\begin{eqnarray}
\label{kin-DY}
&&p_1 = \sqrt{\frac{S}{2}}\, n^*\, ,
\quad p_2 = \sqrt{\frac{S}{2}}\, n\,,\quad
\nonumber\\
&&
n^*_{\mu}=(1/\sqrt{2},\,{\bf 0}_T,\,1/\sqrt{2}), \quad n_{\mu}=(1/\sqrt{2},\,{\bf 0}_T,\,-1/\sqrt{2})\, .
\end{eqnarray}
Accordingly, the quark and gluon momenta $k_1$ and $\ell$ lie
along the plus dominant direction while the gluon momentum $k_2$ -- along the minus direction.
The final on-shell photon and quark(anti-quark) momenta can be presented as
\begin{eqnarray}
\label{Photon-Quark}
q = y_B\, \sqrt{\frac{S}{2}}\, n - \frac{q_\perp^2}{y_B \sqrt{2 S}}\, n^* + q_\perp\,,
\quad
k = x_B\, \sqrt{\frac{S}{2}}\, n^* - \frac{k_\perp^2}{x_B \sqrt{2 S}}\, n + k_\perp\,.
\end{eqnarray}
And, the Mandelstam variables for the process and subprocess are defined as
\begin{eqnarray}
\label{MandVar}
&&S=(p_1+p_2)^2, \quad T=(p_1-q)^2,\quad U=(q-p_2)^2,
\nonumber\\
&& \hat s =(x_1p_1 + yp_2)^2 = x_1 y S, \quad \hat t=(x_1p_1 -q)^2=x_1 T, \quad
\hat u =(q-yp_2)^2 = yU.
\end{eqnarray}
The amplitude (or the hadron tensor) of (\ref{process}) constructed by the contributions from
(i) the leading (LO) diagrams: two diagrams with a radiation of the photon before (${\cal A}^{{\rm LO}}_1$)
and after (${\cal A}^{{\rm LO}}_2$)
the quark-gluon vertex with the gluon going to the lower blob, see the right side of Fig.~\ref{Fig-DirPhot};
(ii) the next-to-leading order (NLO) diagrams: eight diagrams constructed from
the LO diagrams by insertion of all possible radiations
of the additional gluon which together with the quark goes to the upper blob,
see the left side of Fig.~\ref{Fig-DirPhot}.
So, we have the hadron tensor related to the corresponding asymmetry:
\begin{eqnarray}
\label{SSA}
d\sigma^\uparrow - d\sigma^\downarrow \sim {\cal W}=
\sum\limits_{i=1}^{2}\sum\limits_{j=1}^{8}
{\cal A}^{{\rm LO}}_i \ast {\cal B}^{{\rm NLO}}_j\,.
\end{eqnarray}
Here, we will mainly discuss the hadron tensor rather than the asymmetry
itself. So, the hadron tensor as an interference between the LO and NLO diagrams,
${\cal A}^{{\rm LO}}_i \ast {\cal B}^{{\rm NLO}}_j$, can be presented by Fig.~\ref{Fig-DirPhot}
where the upper blob determines the matrix element of the twist-3 quark-gluon operator
while the lower blob -- the matrix element of the twist-2 gluon operator related to the
unpolarized gluon distribution.

\section{Factorization procedure}

The collinear factorization being our main tool, let us outline the main steps of 
the factorization procedure.
It contains 
(i) the decomposition of loop integration momenta around the corresponding dominant direction:
$k_i = x_i p + (k_i\cdot p)n + k_T$
within the certain light cone basis formed by the vectors $p$ and $n$ (in our case, $n^*$ and $n$);
(ii) the replacement:
$d^4 k_i \Longrightarrow d^4 k_i \,dx_i \delta(x_i-k_i\cdot n)$
that introduces the fractions with the appropriated spectral properties;
(iii) the decomposition of the corresponding propagator products around the dominant direction:
\begin{eqnarray}
H(k) = H(xp) + \frac{\partial H(k)}{\partial k_\rho} \biggl|_{k=xp}\biggr. \, k^T_\rho + \ldots \,;
\nonumber
\end{eqnarray}
(iv) the use of the collinear Ward identity, if it requests by the needed approximation:
\begin{eqnarray}
\frac{\partial H(k)}{\partial k_\rho} = H_{\rho}(k,\,k) \, ;
\nonumber
\end{eqnarray}
(v) the Fierz decomposition for $\psi_\alpha (z) \, \bar\psi_\beta(0)$ in
the corresponding space up to the needed projections.
Notice that, for our purposes, it is enough to be limited by the first order of decomposition in the third item.
As a result of this procedure, we should reach the factorized form for the considered subject:
\begin{eqnarray}
\label{Fac-DY}
{\rm Hadron} \,\,\, {\rm tensor} = \{ {\rm Hard} \,\,\, {\rm part} \,\,\, {\rm (pQCD)}\} \otimes
\{ {\rm Soft}\,\,\,{\rm part}\,\,\, {\rm (npQCD)} \}\,.
\end{eqnarray}
Usually, both the hard and soft parts, see (\ref{Fac-DY}), are independent of each other,
UV- and IR-renormalizable and, finally, various parton distributions, parametrizing the soft part,
have to manifest the universality property. 
However, the hard and soft parts of the DPP hadron tensor are not fully independent each other \cite{AT-10, Anikin:2015xka}.
Actually, the DY hadron tensor has formally factorized with the mathematical convolution
and the twist-3 function $B^V(x_1,x_2)$ satisfies still the universality condition.
In contrast to the DY-process, the DPP tensor
will include the functions $B^V(x_1,x_2)$ that will not manifest the universality.

\section{QCD gauge invariance of the hadron tensor}

We now dwell on the QCD gauge invariance of the hadron tensor
for the direct photon production (DPP). 
First of all, let us remind that 
having used the {\it contour gauge conception} \cite{Anikin:2015xka, ContourG}, 
one can check that the representation 
\begin{eqnarray}
\label{B-plus}
B^V_{+}(x_1,x_2)= \frac{T(x_1,x_2)}{x_1-x_2 + i\epsilon}\,
\end{eqnarray}
belongs to the gauge defined by $[x,\,-\infty]=1$, while
the representation 
\begin{eqnarray}
\label{B-minus}
B^V_{-}(x_1,x_2)=\frac{T(x_1,x_2)}{x_1-x_2 - i\epsilon}\,
\end{eqnarray}
corresponds to the gauge that defined by $[+\infty,\, x]=1$. 
In both (\ref{B-plus}) and (\ref{B-minus}), the function $T(x_1,x_2)$ related to the 
following prametrization:
\begin{eqnarray}
\label{parT}
&&\langle p_1, S^T | \bar\psi(\lambda_1 \tilde n)\, \gamma^+ \,
\tilde n_\nu G^{\nu\alpha}_T(\lambda_2\tilde n) \,\psi(0)
|S^T, p_1 \rangle=
\nonumber\\
&&
\varepsilon^{\alpha + S^T -}\,(p_1p_2)\,
\int dx_1 dx_2 \, e^{i x_1\lambda_1+ i(x_2 - x_1)\lambda_2} \, T(x_1,x_2)\,.
\end{eqnarray}
Roughly speaking, it resembles the case where two different vectors have the same projection on the
certain direction. In this sense, the usual axil gauge $A^+=0$ can be understood as a ``projection" which
corresponds to two different ``vectors" represented by two different contour gauges.

\begin{figure}[t]
\centerline{
\includegraphics[width=0.5\textwidth]{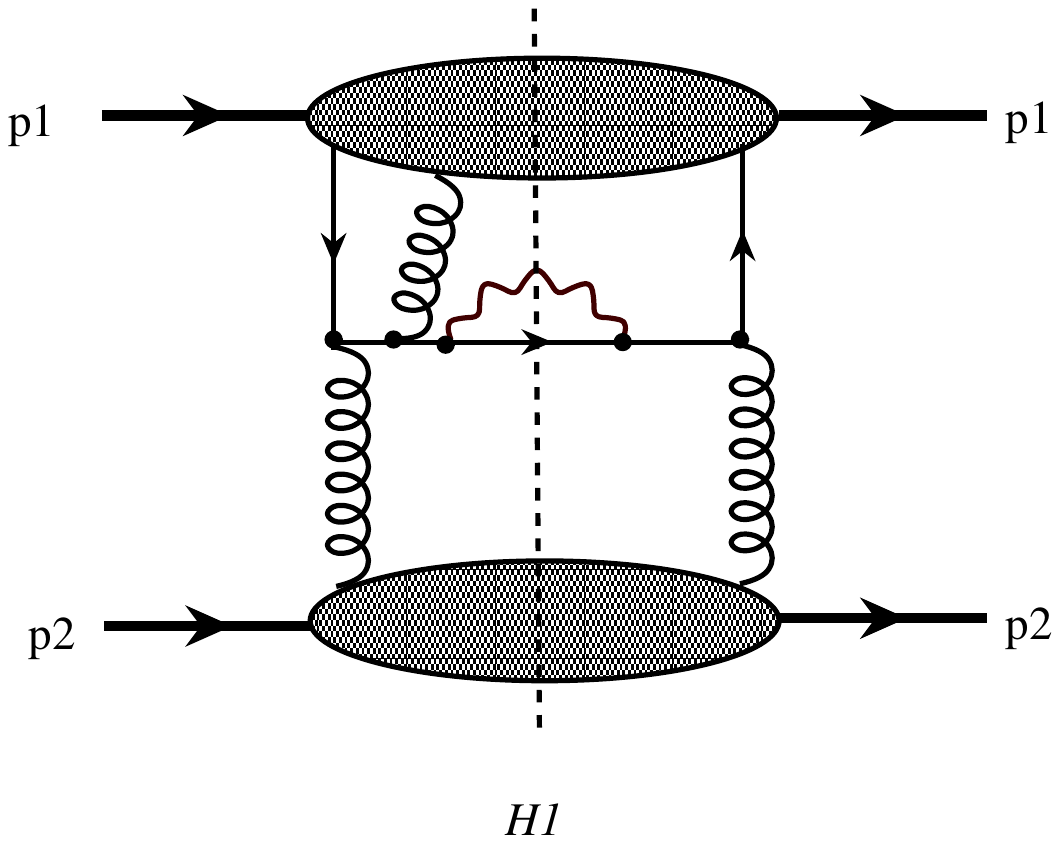}
\hspace{-4.cm}\includegraphics[width=0.5\textwidth]{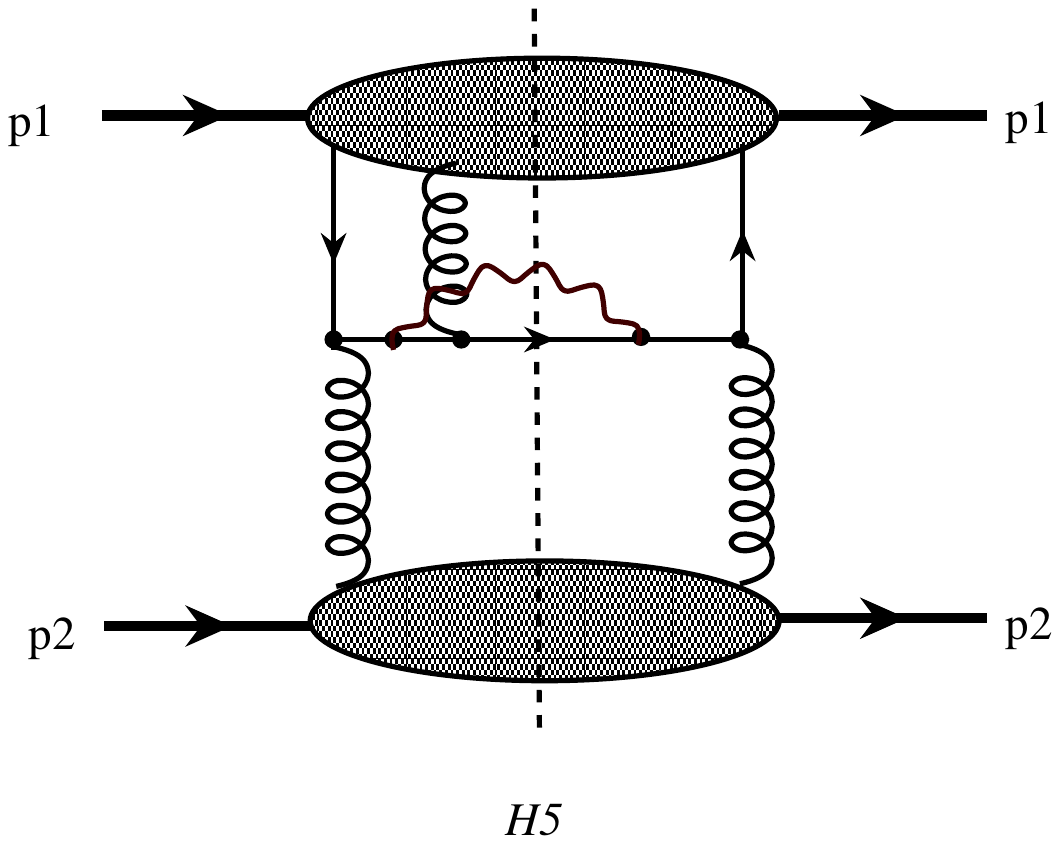}
\hspace{-4.cm}\includegraphics[width=0.5\textwidth]{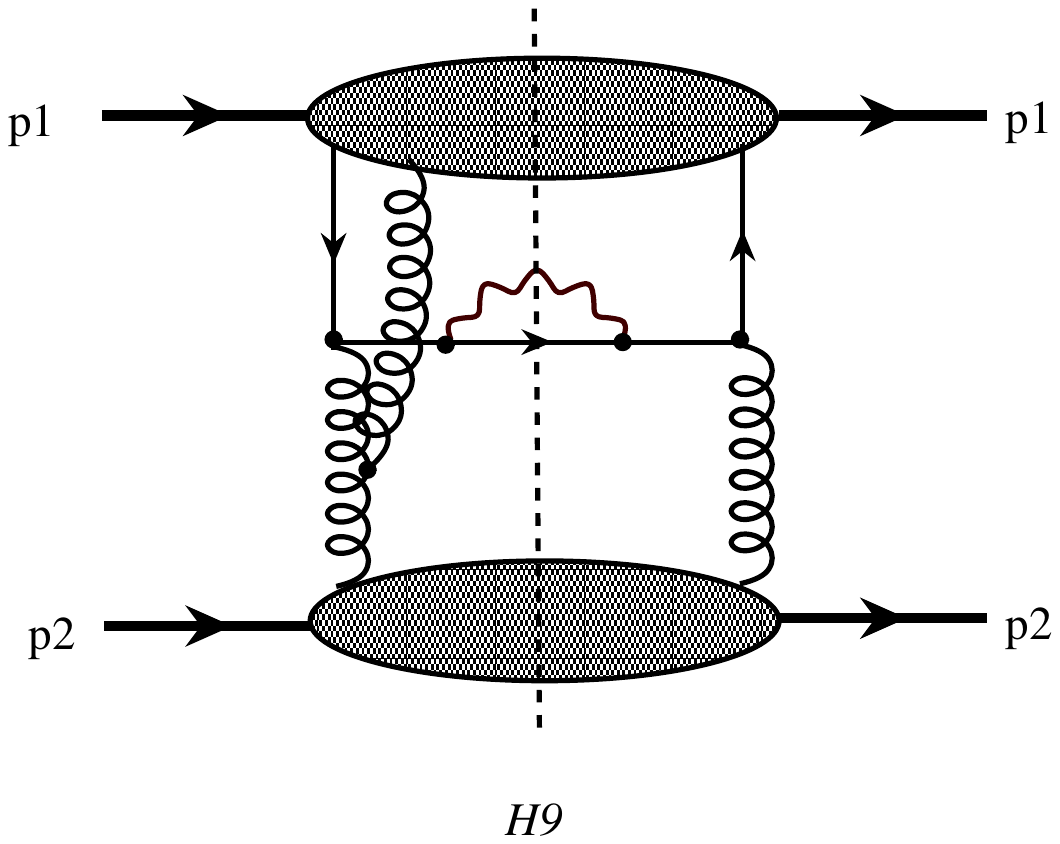}
}
\vspace{-7.5cm}
\centerline{\includegraphics[width=0.5\textwidth]{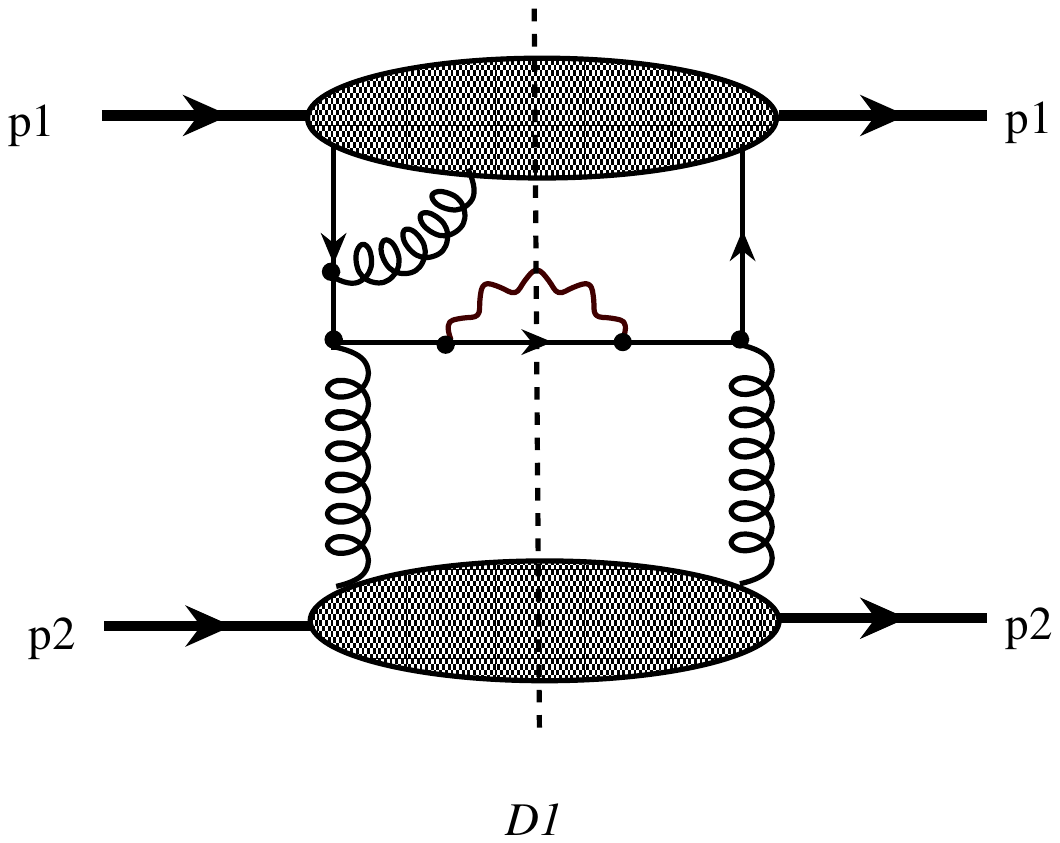}
}
\vspace{-7.cm}
\caption{The typical Feynman diagrams to check the QCD gauge invariance.}
\label{Fig-All}
\end{figure}

Further, to check this invariance,
we have to consider four typical diagrams H1, H5, D1 and H9, depicted in Fig.\ref{Fig-All}, that 
correspond to the certain $\xi$-process (see, \cite{BogoShir}). Notice that, for the QCD gauge invariance, we have to assume
that all charged particles are on its mass-shells. That is, we will deal with only the physical gluons.

To write down the Ward identity, we need to replace the gluon transverse polarization $\epsilon^T_\alpha$
on the gluon longitudinal momentum $\ell^L_\alpha$ in the quark-gluon correlator:
\begin{eqnarray}
\label{qg-col-Rep}
&&\bar\Phi_\perp^{[\gamma^+],\rho}(k_1,\ell)=
- \int (d^4\eta_1)\,
 e^{-ik_1\eta_1}\, \epsilon_T^\rho\,
\langle p_1, S^T| \bar\psi(0)\gamma^+ \psi(\eta_1)\, a^+(\ell)| S^T, p_1\rangle
\stackrel{\epsilon^T\rightarrow \ell^L}{\Longrightarrow}
\nonumber\\
&&- \int (d^4\eta_1)\,
 e^{-ik_1\eta_1}\, \ell_L^\rho\,
\langle p_1, S^T| \bar\psi(0)\gamma^+ \psi(\eta_1)\, a^+(\ell)| S^T, p_1\rangle\,.
\end{eqnarray}
Here, $a^+(\ell)$ stands for the gluon creation operator and the summation over the intermediate states
are not shown explicitly. Notice that the parametrization of this correlator
through $B^V$-function leaves with no any changes in the form.

Consider now the contribution of the H1-diagram, depicted in Fig.\ref{Fig-All}, to the hadron tensor.
Before going further, it is instructive to begin with
the gluon loop integration corresponding to the mentioned diagram, we have
\begin{eqnarray}
\label{gluon-loop-1}
\int (d^4\ell) S(\ell+k+q) \hat\ell_L \langle ... a^+(\ell) ...\rangle\,,
\end{eqnarray}
where we do not write explicitly the irrelevant, at the moment, operators (cf. (\ref{qg-col-Rep})).
After factorization, we obtain
\begin{eqnarray}
\label{gluon-loop-2}
&&\int dx_2 \int (d^4\ell) \delta(x_2-x_1-\ell n) S(\ell+k+q) \hat\ell_L \langle ... a^+(\ell) ...\rangle=
\nonumber\\
&&\int dx_2  S(x_2 p_1+yp_2) \, (x_2-x_1)\hat p_1 \,
\int (d^4\ell) \delta(x_2-x_1-\ell n) \langle ... a^+(\ell) ...\rangle
\,,
\end{eqnarray}
where we decomposed the hard part around the dominant direction  and
used $\ell_L=(x_2-x_1)p_1$, which is actually dictated by the $\gamma$-structure,
together with the momentum conservation. With these, we have the following expression:
\begin{eqnarray}
\label{QCD-WI-1}
&&\overline{{\cal W}}({\rm diag.H1})=
-2 \,\int dx_1\, dy \, \delta^{(4)}(x_1 p_1+y p_2-k-q) \,
{\cal F}^g(y)\, {\rm C}_2 \times
\nonumber\\
&&\bar v(k)\hat\varepsilon \, \frac{\gamma^+}{2x_1 p_1 + i\epsilon}
 \gamma^- \,
\int dx_2\,  \frac{(x_2-x_1) \gamma^+\gamma^-}{2x_2 + i\epsilon}
 \frac{\gamma^+}{2x_1 p_1 + i\epsilon} \hat\varepsilon^*  v(k)\times\,
\nonumber\\
&&
\Big\{
 (-)\int (d\lambda_1)\,
 e^{-ix_1\lambda_1}\,
\langle p_1, S^T| \bar\psi(0)\gamma^+ \psi(\lambda_1 n)\,
\int (d^4\ell) \delta(x_2-x_1-\ell n)\, a^+(\ell)| S^T, p_1\rangle
\Big\}\,.
\end{eqnarray}
We can here use the parametrization in the form of
\begin{eqnarray}
\label{B-par-QCD}
&&(-)\int (d\lambda_1)\,
 e^{-ix_1\lambda_1}\,
\langle p_1, S^T| \bar\psi(0)\gamma^+ \psi(\lambda_1 n)\,
\int (d^4\ell) \delta(x_2-x_1-\ell n) a^+(\ell)| S^T, p_1\rangle 
\nonumber\\
&&= B^V(x_1,x_2)\,.
\end{eqnarray}
It is seen, however, that in any case this diagram does not contribute to the Ward identity
after calculation of the imaginary
part owing to the factor $(x_2-x_1)$ in the numerator of (\ref{QCD-WI-1}).

Further, calculation of the H5-diagram, presented in Fig.\ref{Fig-All}, gives us
\begin{eqnarray}
\label{QCD-WI-2}
&&\overline{{\cal W}}({\rm diag.H5})=
\int dx_1\, dy \, \delta^{(4)}(x_1 p_1+y p_2-k-q) \,
{\cal F}^g(y)\, {\rm C}_2 \times
\nonumber\\
&&\bar v(k)\hat\varepsilon \, \frac{\gamma^+}{2x_1 p_1 + i\epsilon}
 \gamma^- \,  \hat\varepsilon^*\,
\int dx_2\,  \frac{\gamma^+\gamma^-\gamma^+}{2x_2p_1 + i\epsilon}\, v(k)
\, B^V(x_1,x_2)\,,
\end{eqnarray}
while the contribution of the D1-diagram in Fig.\ref{Fig-All} takes the form
\begin{eqnarray}
\label{QCD-WI-3}
&&\overline{{\cal W}}({\rm diag.D1})=
- \int dx_1\, dy \, \delta^{(4)}(x_1 p_1+y p_2-k-q) \,
{\cal F}^g(y)\, {\rm C}_1 \times
\nonumber\\
&&\bar v(k)\hat\varepsilon \, \frac{\gamma^+}{2x_1 p_1 + i\epsilon}
 \gamma^- \gamma^+ \gamma^-\,
\frac{\gamma^+}{2x_1p_1 + i\epsilon}\,  \hat\varepsilon^*\, v(k)
\int dx_2\, \, B^V(x_1,x_2)\,.
\end{eqnarray}
And, finally, the contribution of the H9-diagram with the three-gluon vertex, see Fig.\ref{Fig-All},
reads
\begin{eqnarray}
\label{QCD-WI-4}
&&\overline{{\cal W}}({\rm diag.H9})=
-4\,i\, \int dx_1\, dy \, \delta^{(4)}(x_1 p_1+y p_2-k-q) \,
{\cal F}^g(y)\, {\rm C}_3 \times
\nonumber\\
&&\bar v(k)\hat\varepsilon \, \frac{\gamma^+}{2x_1 p_1 + i\epsilon}
 \gamma^- \, \frac{\gamma^+}{2x_1p_1 + i\epsilon}\, \hat\varepsilon^*\, v(k)
\int dx_2\, \frac{x_2-x_1}{2(x_2-x_1)+i\epsilon} \, B^V(x_1,x_2)\,.
\end{eqnarray}
We now turn to the contour gauge. Based on the discussions in \cite{Anikin:2015xka}, 
even at glance, we are able  to anticipate the corresponding prescriptions for $B^V$-functions in
(\ref{QCD-WI-2})--(\ref{QCD-WI-4}).
Indeed, the H5-diagram in Fig.\ref{Fig-All} corresponds to the final state interaction
and, therefore, the function $B^V_-$
should appear here, while the D1- and H9-diagrams in Fig.\ref{Fig-All} -- to the
initial state interaction which leads to the function $B^V_+$.
Performing the explicit calculations (as we did for the DY-process \cite{AT-10}),
we can obtain the same conclusion by restoring the Wilson lines in the quark-gluon correlators
of the mentioned diagrams.
That is, the Wilson line $[+\infty^-,\, z^-]$ will enter in the hadron tensor represented by
the H5-diagram in Fig.\ref{Fig-All};
the Wilson line $[z^-,\,-\infty^- ]$ will stand in the hadron tensor represented by
the D1- and H9-diagrams in Fig.\ref{Fig-All}.

So that, we sum all contributions and get the following final expression:
\begin{eqnarray}
\label{QCD-WI-final}
\sum\limits_{{\rm N}}\overline{{\cal W}}({\rm diag. N})&=&
\frac{{\rm C}_2}{8 x_1} \gamma^+ \gamma^- \gamma^+ \gamma^- \gamma^+
\int dx_2 \frac{B^V_-(x_1,x_2)}{x_2} +
\frac{{\rm C}_1}{8 x_1^2} \gamma^+ \gamma^- \gamma^+ \gamma^- \gamma^+
\int dx_2 B^V_+(x_1,x_2) +
\nonumber\\
&&
\frac{i{\rm C}_3}{4 x_1^2} \gamma^+ \gamma^- \gamma^+
\int dx_2 \frac{(x_2-x_1) B^V_+(x_1,x_2)}{x_2-x_1+i\epsilon}\,
\end{eqnarray}
where the function $B^V_{\pm}$ are given by (\ref{B-plus}) and (\ref{B-minus}).

We now calculate the imaginary part and, ultimately, derive the QCD Ward identity in the form
\begin{eqnarray}
\label{QCD-WI}
{\rm C}_2 - {\rm C}_1 - i{\rm C}_3 =
- [t^a, t^b]\, t^b\, t^a + i f^{abc} t^c\, t^b\, t^a  \equiv 0\,.
\end{eqnarray}
We want to stress that the identity (\ref{QCD-WI}) takes place provided only the
presence of the different complex prescriptions in gluonic poles dictated by the final or
initial state interactions:
\begin{eqnarray}
\label{QCD-illust}
  \begin{array}{rcl}
\hspace{-2cm}{\rm \bf FSI}&\Rightarrow& \frac{1}{\ell^+ + i\epsilon} \Rightarrow {\rm gauge}\,\,\,[+\infty^-,\, z^- ]=1
\Rightarrow \frac{T(x_1,x_2)}{x_1-x_2 - i\epsilon}\,\\
\\
\hspace{-2cm}{\rm \bf ISI}&\Rightarrow& \frac{1}{-\ell^+ + i\epsilon} \Rightarrow {\rm gauge}\,\,\,[z^-,\, -\infty^-]=1
\Rightarrow \frac{T(x_1,x_2)}{x_1-x_2 +i\epsilon}\,\\
  \end{array}\,\,\, \Bigg\}
\Rightarrow{\rm \bf QCD}\,\,\, {\rm \bf GI}\,.
\end{eqnarray}
We emphasize the principle differences between the considered case and the proof
of the QCD gauge invariance for the perturbative Compton scattering amplitude with the physical gluons in the
initial and final states. The latter does not need any external condition, like the presence of gluon poles.

Thus, the situation which we discuss is  again absolutely similar to that one which was described in
\cite{Braun} for the dijet production. From (\ref{QCD-illust}), it is seen that
the different diagrams correspond
to the different contour gauges and, consequently, to different the
functions $B^V_{\pm}$ that parametrize the hadronic
matrix element forming the soft part. In this context, we also have
a soft breaking of factorization because, first, it spoils
the universality principle and, second, the gluonic pole prescriptions in
the soft part have been traced to the causal prescriptions in
the hard part. Besides, the possible reasons for the 
collinear factorization breaking has been presente in \cite{Anikin:2015xka}.

\section{Conclusions}

We explore the QCD gauge invariance of
the hadron tensor for the direct photon production in two
hadron collision where one of hadrons is transversely polarized.
We argue the effects which lead to the soft breaking of factorization
by inspection of the QCD gauge invariance.
We demonstrate that 
the initial or final state interactions in diagrams define
the different prescriptions in the gluonic poles.
Moreover, the different prescriptions are needed to ensure the QCD gauge invariance.
This situation can be treated as a soft breaking of the
universality condition resulting in factorization breaking.

\subsection*{Acknowledgements}

We would like to thank I.O.~Cherednikov, A.V.~Efremov, D.~Ivanov and
L.~Szymanowski for useful discussions and correspondence.
This work is partly supported by the HL program.

\end{document}